\documentclass[aps,prd,twocolumn,superscriptaddress,showpacs,showkeys]{revtex4}

\usepackage{epsfig,epsf}
\usepackage{amsmath}
\usepackage{amsthm}
\usepackage{amsfonts}
\usepackage{amssymb}
\usepackage{dsfont}

\usepackage{multirow}

\usepackage{slashed}

\usepackage[active]{srcltx}

\def\OMIT#1{}

\newcommand{\be}{\begin{equation}}
\newcommand{\ee}{\end{equation}}

\DeclareMathOperator{\Li}{Li}

\newcommand \vev [1] {\langle{#1}\rangle}

\def\II{\hbox{{1}\kern-.25em\hbox{l}}}
\begin{document}


\title{Finite--$t$ and target mass corrections to deeply virtual Compton scattering}


\date{\today}

\author{V.M.~Braun}
\affiliation{Institut f\"ur Theoretische Physik, Universit\"at
   Regensburg,D-93040 Regensburg, Germany}
\author{A.N.~Manashov}
\affiliation{Institut f\"ur Theoretische Physik, Universit\"at
   Regensburg,D-93040 Regensburg, Germany}
\affiliation{Department of Theoretical Physics,  St.-Petersburg 
University, 199034, St.-Petersburg, Russia}
\author{B.~Pirnay}
\affiliation{Institut f\"ur Theoretische Physik, Universit\"at
   Regensburg,D-93040 Regensburg, Germany}

\date{\today}

\begin{abstract}
  \vspace*{0.3cm}
\noindent We carry out the first complete calculation of kinematic power corrections $\sim
t/Q^2$ and $\sim m^2/Q^2$ to the helicity amplitudes of deeply-virtual Compton scattering.
This result removes an important source of uncertainties in the QCD predictions for
intermediate momentum transfers $Q^2\sim 1-10$~GeV$^2$ that are accessible in the existing
and planned experiments. In particular the finite--$t$ corrections are significant and
must be taken into account in the data analysis.
 \end{abstract}



\maketitle


Deeply Virtual Compton Scattering (DVCS) is the simplest process that gives access to
generalized parton distributions (GPDs) and is receiving a lot of
attention~\cite{Diehl:2003ny,Belitsky:2005qn}. The existing experimental results come from
HERMES and Jefferson Lab (Hall A and CLAS) and many more measurements are planned after
the Jefferson Lab $12$~GeV upgrade and at COMPASS-II at CERN. Since the bulk of the
existing and expected data is for photon virtualities $Q^2<5$~GeV$^2$, corrections of the
type $m^2/Q^2,\,\, t/Q^2$, where $m$  is the target (nucleon) mass and $t=(p'-p)^2$ is
the momentum transfer to the target,
can have significant impact on the data analysis and should be taken into account. The
finite--$t$ corrections are of particular importance if one wants to study the
three-dimensional picture of the proton in longitudinal and transverse
plane~\cite{Burkardt:2002hr}, in which case the $t$--dependence has to be measured in a
sufficiently broad range.

The necessity of taking into account kinematic power corrections to DVCS is widely acknowledged%
~\cite{Belitsky:2005qn,Anikin:2000em,Blumlein:2000cx,Kivel:2000rb,Radyushkin:2000ap,Belitsky:2000vx,Belitsky:2001hz,Geyer:2004bx,Blumlein:2006ia,Blumlein:2008di,Belitsky:2010jw}.
Early attempts to calculate such corrections by analogy to Nachtmann
corrections~\cite{Nachtmann:1973mr} to the structure functions in deep-inelastic
lepton-nucleon scattering produced results that were not gauge invariant and not
translation invariant with respect to the choice of the positions of the electromagnetic
currents. The reason is that in addition to Nachtmann-type contributions related to
subtraction of traces in the leading-twist operators one must take into account their
higher-twist descendants obtained by adding total derivatives: $ \mathcal{O}_1 \sim
\partial^2 \mathcal{O}_{\mu_1\ldots\mu_n}$, and $\mathcal{O}_2 \sim
\partial^{\mu_1}\mathcal{O}_{\mu_1\ldots\mu_n}$, where $\mathcal{O}_{\mu_1\ldots\mu_n}$
are the usual leading-twist operators. The problem arises because matrix elements of the
operator $\mathcal{O}_2$ on free quarks vanish~\cite{Ferrara:1972xq}. Thus in order to
find its \emph{leading-order} coefficient function in the operator product expansion (OPE)
of two electromagnetic currents one is forced to consider either more complicated
(quark-antiquark-gluon) matrix elements, or stay with the quark-antiquark ones but go over
to the next-to-leading order in
$\alpha_s$. 
In both cases the real difficulty is not the calculation of the relevant Feynman diagrams,
but the necessity to separate the contribution of interest from the ``genuine''
quark-gluon twist-four
operators.

This problem was solved in Refs.~\cite{Braun:2011zr,Braun:2011dg} using conformal symmetry
which implies that coefficient functions of ``kinematic'' and ``genuine'' twist-four
operators are mutually orthogonal with  a proper weight
function~\cite{Braun:2009vc,Braun:2012bg}. Using this approach we have calculated in
Ref.~\cite{Braun:2012bg} the finite--$t$ and target-mass corrections to DVCS for the study
case of a scalar target. We verified gauge- and translation-invariance and, most
importantly, found that the structure of kinematic corrections proves to be consistent
with collinear factorization. In this letter we present our final results for the helicity
amplitudes of DVCS to the $1/Q^2$ accuracy for the physically interesting case of the
spin-1/2 (nucleon) target. This result removes one important source of uncertainties in
the QCD predictions for intermediate photon virtualities  that are accessible in the
existing and planned experiments.


The DVCS amplitude $\gamma^*(q)+N(p)\to \gamma(q')+N(p')$ is defined by the matrix element
of the time-ordered product of two electromagnetic currents, sandwiched between the
nucleon states
\begin{eqnarray}
\lefteqn{\hspace*{-1cm}i\int\! d^4 x\int\! d^4 y\, e^{-iqx+iq'y}
\vev{p'|T\{j^{\rm em}_\mu(x)j^{\rm em}_\nu(y)\}|p}=}
\notag\\
&=& (2\pi)^4\delta(p+q-p'-q')\,\mathcal{A}_{\mu\nu}(q,q',p)\,.
\label{Tproduct}
\end{eqnarray}
Introducing the  photon polarization vectors \mbox{$(\varepsilon^{\pm,0} q)=0$},
$(\varepsilon^\pm q')=0$ one can write $\mathcal{A}_{\mu\nu}$ in terms of helicity
amplitudes
\begin{align}
\!\mathcal{A}_{\mu\nu}=&\,\, \varepsilon^+_{\mu} \varepsilon^-_{\nu} \mathcal{A}^{++}
+\varepsilon^-_{\mu} \varepsilon^+_{\nu} \mathcal{A}^{--}
+\varepsilon^0_{\mu} \varepsilon^-_{\nu} \mathcal{A}^{0+}
\notag\\
&+
\varepsilon^0_{\mu} \varepsilon^+_{\nu} \mathcal{A}^{0-}\!
+\!\varepsilon^+_{\mu} \varepsilon^+_{\nu} \mathcal{A}^{+-}\!
+\!\varepsilon^-_{\mu} \varepsilon^-_{\nu} \mathcal{A}^{-+}\!+\!q'_\nu\mathcal{A}_\mu^{(3)}\!.
\label{Amunu}
\end{align}
The last term $\sim q'_\nu $ is of no interest as it does not contribute to any
observable.

The helicity-conserving amplitudes are the leading ones in the scaling limit,
$
A^{\pm\pm} \sim \mathcal{O}(Q^0)$, and the helicity-flip amplitudes are power-suppressed:
$
A^{0\pm}\sim \mathcal{O}(Q^{-1})$,
$
A^{\pm\mp}\sim \mathcal{O}(Q^{-2})$. Thus in order to calculate physical observables to
the $1/Q^2$ accuracy one has to take into account $1/Q^2$ corrections to $A^{++}$ and
$A^{--}$, whereas for the helicity-flip amplitudes the leading power accuracy is
sufficient.

The definition of helicity amplitudes depends on a reference frame. We use the photon
momenta, $q$ and  $q'$, to define a longitudinal plane spanned by the two light-like
vectors
\begin{equation}
n=q'\,, \qquad \tilde n=-q+(1-\tau)\, q'\,,
\end{equation}
where $\tau= t/(Q^2+t)$, $Q^2=-q^2$. For this choice the momentum transfer to the target
$\Delta = p'-p= q-q'$, $ t=\Delta^2$ is purely longitudinal and the target (proton)
momenta have a nonzero transverse component
\begin{align}\label{Pperp}
| P_\perp |^2=-m^2-\frac{t}{4}\frac{1-\xi^2}{\xi^2}\sim \mathcal{O}(Q^0)\,,
\end{align}
where $P=(p+p')/2$ and the skewedness parameter is defined as
$\xi=-(\Delta\cdot q') /(2(P\cdot q'))$\,.

The condition $| P_\perp |^2 > 0$ translates to the lower bound $|t|
> |t_{\rm min}| = 4m^2\xi^2/(1-\xi^2)$, cf.~\cite{Belitsky:2005qn}.

We choose the polarization vectors as follows~\cite{Braun:2012bg}
\begin{align}
\varepsilon^0_\mu=&-\left(q_\mu-q'_\mu {q^2}/{(qq')}\right)/{\sqrt{-q^2}}\,,\notag\\
\varepsilon^\pm_\mu=&(P^\perp_\mu\pm i \bar P^\perp_\mu)/\sqrt{2}\,,
\end{align}
where $P^\perp_\mu=g_{\mu\nu}^\perp P^\nu$, $\bar P^\perp_\mu=\epsilon_{\mu\nu}^\perp
P^\nu$ and
\begin{align}
g_{\mu\nu}^\perp=&g_{\mu\nu}-(q_\mu q'_\nu+q'_\mu q_\nu)/(qq')+{q'_\mu}q'_\nu\,{q^2}/(qq')^2\,,
\notag\\
\epsilon_{\mu\nu}^\perp=&\epsilon_{\mu\nu\alpha\beta}{q^\alpha q'^\beta}/(qq')\,.
\end{align}

Each helicity amplitude involves the sum over quark flavors, $\mathcal{A}=\sum e_q^2
\mathcal{A}_q$, where $e_q$ is the quark electromagnetic charge, and is written in terms
of the leading-twist GPDs $H^q,E^q,\widetilde H^q,\widetilde E^q$. For the GPD definitions
we follow Ref.~\cite{Diehl:2003ny}.

The calculation is similar to the case of the scalar target considered in
Ref.~\cite{Braun:2012bg} so that in this letter we only present the final expressions.
Note that the electromagnetic gauge invariance is guaranteed to twist-four accuracy
already on the operator level and is embedded in the definition of helicity amplitudes.
The translation invariance (independence on the shift of the positions of the
electromagnetic currents in Eq.~(\ref{Tproduct}): $x\to x+\delta, y\to y+\delta$) is
nontrivial and provides a strong check of the calculation, see Ref.~\cite{Braun:2012bg}.
The results can conveniently be written in terms of the vector and axial-vector bispinors
\begin{align}\label{Dirac}
v^\mu=\bar u(p')\gamma^\mu u(p)\,, &&a^\mu=\bar u(p')\gamma^\mu \gamma_5u(p)\,.
\end{align}
We define
%
$ v_\perp^\pm=(v\cdot\varepsilon^\pm)\,,\quad a_\perp^\pm=(a\cdot\varepsilon^\pm)\,,\quad
P_\perp^\pm=(P\cdot \varepsilon^\pm)\,. $
%
Although $P_\perp^+=P_\perp^-=|P_\perp|/\sqrt{2}$ we prefer this notation to keep trace of
the  polarization vectors. We also use a shorthand notation
\begin{align}
X^q(x,\xi,t)=H^q(x,\xi,t)+E^q(x,\xi,t)\,
\end{align}
and rewrite the helicity-conserving amplitudes in terms of the vector- and axial-vector
invariant functions as
\begin{align}
\frac12(\mathcal{A}_q^{++}\!+\!\mathcal{A}_q^{--})=&
\frac{(vP)}{2m^2}\, \mathbb{V}^q_1+\frac{(v q')}{(qq')}\, \mathbb{V}^q_2\,,
\label{V1}\\
\frac12(\mathcal{A}_q^{++}\!-\!\mathcal{A}_q^{--})=&\frac{(a\Delta)}{4m^2}\, \mathbb{A}^q_1
+\frac{(a q')}{(qq')}\, \mathbb{A}^q_2\,.
\end{align}
The following expressions for $\mathbb{V}^q_k$, $\mathbb{A}^q_k$ present our main result:
\begin{widetext}
\begin{subequations}
\begin{align}
\mathbb{V}^q_1
=&\Big(1-\frac{t}{2Q^2}\Big) E^q\otimes{C}_0^-
+\frac{t}{Q^2}E^q\otimes\mathcal{ C}_1^-
-\frac2{Q^2}\Big(\frac{t}\xi+2|P_\perp|^2\xi^2\partial_\xi\Big)\xi^2\partial_\xi
  E^q\otimes{C}_2^-
+\frac{8m^2}{Q^2}\xi^2\partial_\xi \xi\, X^q\otimes {C}_2^-\,,
\label{Vq1}\\
\mathbb{V}^q_2
=&\Big(1-\frac{t}{2Q^2}\Big)\xi\, X^q\otimes {C}_0^-
+\frac{t}{Q^2}\xi\,X^q\otimes\mathcal{ C}_1^-
-\frac4{Q^2}\biggl[\Big(|P_\perp|^2\xi^2\partial_\xi+\frac{t}{\xi}\Big)\xi^2\partial_\xi-\frac{t}2\biggr]
 \xi\,X^q\otimes  {C}_2^-\,,
\label{Vq2}\\
\mathbb{A}^q_1
=& \Big(1-\frac{t}{2Q^2}\Big) \xi\,\widetilde E^q\otimes{C}_0^++\frac{t}{Q^2}
\xi\,\widetilde E^q\otimes{C}_1^+
-\frac2{Q^2}\Big(\frac{t}\xi+2|P_\perp|^2\xi^2\partial_\xi\Big)\xi^2\partial_\xi
 \xi\, \widetilde E^q\otimes{C}_2^++\frac{8m^2}{Q^2}\xi^2\partial_\xi \widetilde
H^q\otimes{C}_2^+\,,
\label{Aq1}\\
\mathbb{A}^q_2
=& \Big(1-\frac{t}{2Q^2}\Big)\xi\,\widetilde H^q \otimes{C}_0^+
 +\frac{t}{Q^2}\xi\,\widetilde H^q\otimes {C}_1^+
-\frac4{Q^2}\biggl[\Big(|P_\perp|^2\xi^2\partial_\xi+\frac{t}{\xi}\Big)\xi^2\partial_\xi-\frac{t}2\biggr]
\xi\,\widetilde H^q\otimes{C}_2^+\,.
\label{Aq2}
\end{align}
\end{subequations}
\end{widetext}
In addition, for the helicity-flip amplitudes we obtain
\begin{align}
\mathcal{A}_q^{0,\pm}=&\frac2{Q}
\biggl\{
\left(v_\perp^\pm-4P_\perp^\pm \frac{(vq')}{Q^2}
\xi^2\partial_\xi\right)\xi\, X^q\otimes{C}_1^-
\notag\\
&\hspace*{0.5cm}
\pm\left(a_\perp^\pm-4P_\perp^\pm \frac{(aq')}{Q^2}
\xi^2\partial_\xi\right)\xi\, \widetilde{H}^q\otimes{C}^+_1
\notag\\
&\hspace*{0.5cm}
+P_\perp^\pm\frac{(vP)}{m^2}\xi^2\partial_\xi\, E^q\otimes {C}_1^-
\notag\\
&\hspace*{0.5cm}
\pm P_\perp^\pm\frac{(a\Delta)}{2m^2}\xi^2\partial_\xi\,\xi
\,\widetilde E^q\otimes{C}_1^+\biggr\}
\label{A0}
\end{align}
and
\begin{align}
\label{Aflip}
\mathcal{A}_q^{\mp\pm}=&
-\frac{8P_\perp^\pm}{Q^2} \biggl\{\Big(v_\perp^\pm-2P_\perp^\pm\frac{(vq')}{Q^2}\,\xi^2\partial_\xi\Big)\xi^2\partial_\xi
\,X^q\otimes{[x {C}_1^-]}
\notag\\
&\hspace*{0.5cm}
\pm\Big(a_\perp^\pm-2P_\perp^\pm
\frac{(aq')}{Q^2}\xi^2\partial_\xi\Big)\xi^2\partial_\xi
\,\xi\,\widetilde H^q\otimes{C}_1^+
\notag\\
&\hspace*{0.5cm}
+P_\perp^\pm\frac{(vP)}{2m^2}\xi^3\partial_\xi^2
\,E^q\otimes{[x C_1^-]}
\notag\\
&\hspace*{0.5cm}
\mp P_\perp^\pm\frac{(a\Delta)}{4m^2}\xi^3
\partial_\xi^2\xi^2
\widetilde E^q\otimes{C}_1^+
\biggr\}.
\end{align}
In all cases $\partial_\xi =\partial/\partial\xi$ and the notation  $F\otimes C$ stands
for the convolution of a GPD $F$ with a coefficient function $C$:
$$
F\otimes C\equiv\int dx\, F(x,\xi,t)\, C(x,\xi)\,.
$$
The coefficient functions ${C}_k^\pm(x,\xi)$ are  given by the following expressions:
%
\begin{align}\label{Coefff}
{C}_0^\pm(x,\xi)=&\frac{1}{\xi+x-i\epsilon}\pm\frac{1}{\xi-x-i\epsilon}\,,
\notag\\
{C}^\pm_1(x,\xi)=&\frac{1}{x-\xi}\ln\Big(\frac{\xi+x}{2\xi}-i\epsilon\Big)\pm (x\leftrightarrow -x)\,,
\notag\\
{C}_2^\pm(x,\xi)=&
\biggl\{\frac{1}{\xi+x}\Big[\Li_2\Big(\frac{\xi-x}{2\xi}+i\epsilon\Big)-\Li_2(1)\Big]
\notag\\
&\pm
(x\leftrightarrow -x)\biggr\}+\frac12 {C}_1^\pm(x,\xi)\,.
\end{align}
Note that ${C}_0^\pm$ have simple poles at $x=\pm \xi$ whereas ${C}_{1,2}^\pm$ have a
milder (logarithmic) singularity at the same points. This ensures that the kinematic power
corrections are factorizable, at least to the leading order in $\alpha_s$. In the DVCS
kinematics
\begin{align}
(vq') \sim (aq') = \mathcal{O}(Q^2)\,,&&
(vP) \sim (a\Delta) = \mathcal{O}(Q^0)
\end{align}
so that the helicity-conserving amplitudes (\ref{Vq1})~--~(\ref{Aq2}) include leading
contributions $\mathcal{O}(1/Q^0)$ and the corrections $\mathcal{O}(1/Q^2)$, whereas all
terms in Eqs.~(\ref{A0}) and (\ref{Aflip}) are of the order $\mathcal{O}(1/Q)$ and
$\mathcal{O}(1/Q^2)$, respectively, as expected.

Our results for the DVCS helicity amplitudes have a similar structure to those in
Ref.~\cite{Braun:2012bg} for the scalar target~\cite{correction}. 
The main difference is the appearance of a
large target mass correction to the contribution of the GPD $E^q$ ($\widetilde{E}^q$) that
involves $X^q$ ($\widetilde{H}^q$), cf. the last term in the first line of Eq.~(\ref{Vq1})
(Eq.~(\ref{Aq1})).

It has become customary to parametrize the DVCS amplitude (\ref{Amunu}) in terms of the
so-called Compton form factors (CFFs) $\mathcal{H}$ and
$\mathcal{E}$~\cite{Belitsky:2005qn}
\begin{align}
\mathcal{A}_{\mu\nu}= - \frac{g_{\mu\nu}^\perp}{2(Pq')}\left[(\bar u \slashed{q}' u)\,\mathcal{H}
+\frac{i}{2m}(\bar u\sigma^{\mu\nu}q'_\mu\Delta_\nu u)\,\mathcal{E}\right] +\ldots
\end{align}
In our notation
\begin{align}
\mathcal{H}=\mathbb{V}_1-{\mathbb{V}_2}/\xi\,, &&\mathcal{E}=-\mathbb{V}_1\,,
\end{align}
where $\mathbb{V}_k = \sum_q e^2_q \mathbb{V}^q_k$.

A detailed study of the numerical impact of the kinematic corrections on different DVCS
observables goes beyond the tasks of this letter. For orientation, we have calculated the
corrections to the imaginary parts of the CFFs $\mathcal{H}$ and $\mathcal{E}$ which
involve the GPDs in the DGLAP region only. To this end we use the GPD model of
Refs.~\cite{Hyde:2011ke,Guidal:2004nd}:
\begin{align}
\left\{
\begin{matrix}H^q\\E^q\end{matrix}
\right\}(x,\xi,t)=\int d\beta\int d\alpha\,\delta(x-\beta-\xi\alpha)f_q(\beta,\alpha,t)\,,
\nonumber
\end{align}
where the double distribution $f_q$ is written as
\begin{align}
f_q(\beta,\alpha,t)=h(\beta,\alpha)|\beta|^{-\alpha't}q(\beta)
\left\{\begin{matrix}1\\
\kappa_q (1-\beta)^{\eta_q}/A_q
\end{matrix}\right.\,.
\nonumber
\end{align}
Here $q(\beta)$ is the MRST2002 NNLO valence $u$- and $d$-quark
distribution~\cite{Martin:2002dr}) and the profile function $h$ is given by the following
expression:
\begin{align}
h(\beta,\alpha)=\frac34{((1-|\beta|)^2-\alpha^2})/{(1-|\beta|)^3}\,,
\nonumber
\end{align}
where $k_u\simeq 1.7$ and $k_d\simeq-2.0$ are the anomalous magnetic moments, $\eta_u
\simeq 1.7$ and $\eta_d\simeq 0.57$~\cite{Guidal:2004nd}, and $A_q=\int
d\beta\,(1-\beta)^{\eta_q} q(\beta)$.
\begin{figure}[t]
{}\hspace*{3mm}\includegraphics[width=0.865\columnwidth]{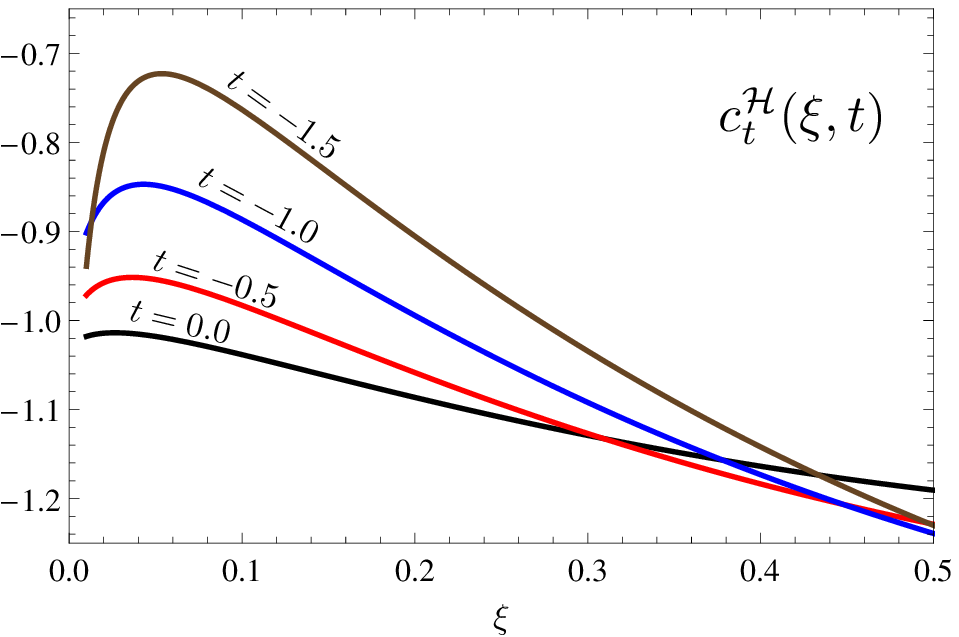}\\[2mm]
{}\hspace*{3mm}\includegraphics[width=0.865\columnwidth]{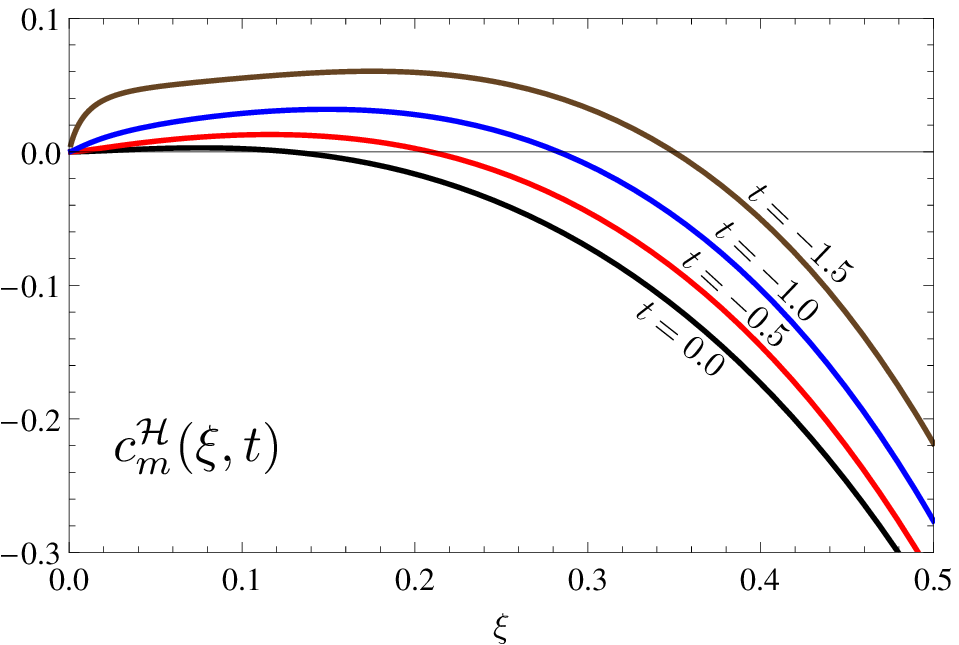}
\caption{The coefficients $c_t^\mathcal{H}(\xi,t)$ (upper panel) and $c_{m}^\mathcal{H}(\xi,t)$
(lower panel) for  different values of $t$ (in $\text{GeV}^2$).}
\label{figure1}
\end{figure}
\begin{figure}[t]
{}\hspace*{2mm}\includegraphics[width=0.865\columnwidth]{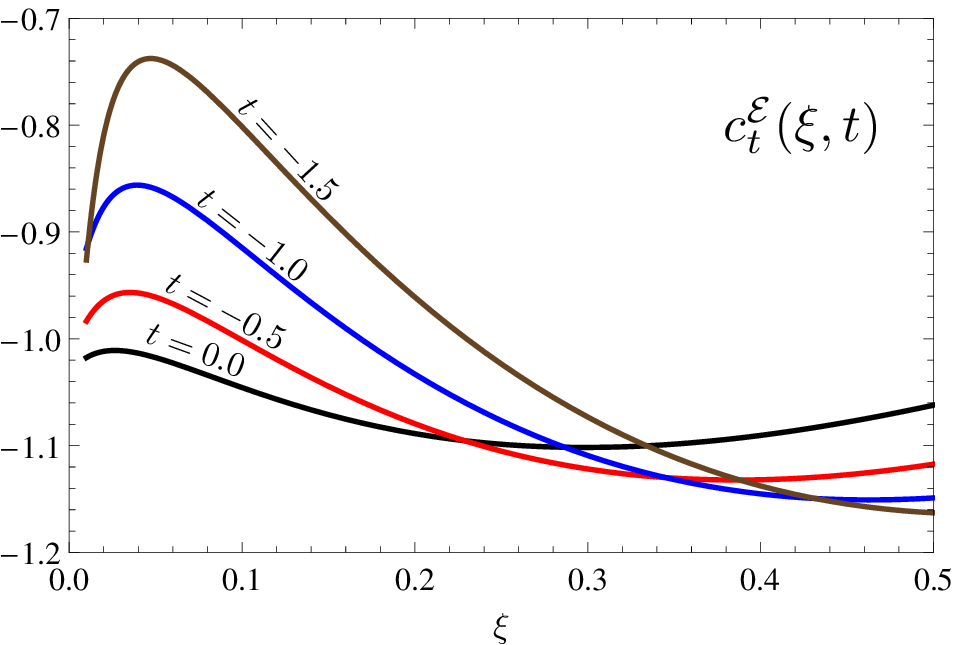}\\[2mm]
{}\hspace*{5mm}\includegraphics[width=0.85\columnwidth]{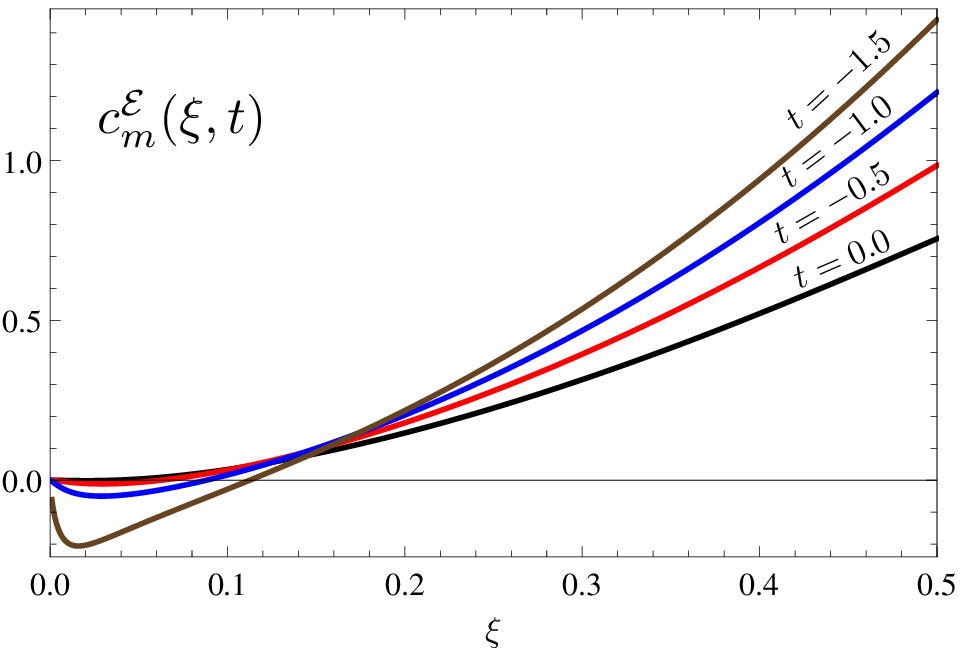}
\caption{The coefficients $c_t^\mathcal{E}(\xi,t)$ (upper panel) and $c_{m}^\mathcal{E}(\xi,t)$
(lower panel) for  different values of $t$ (in $\text{GeV}^2$).}
\label{figure2}
\end{figure}
We consider the following ratios
\begin{align}
\frac{\text{Im} \mathcal{F}-\text{Im} \mathcal{F}^{LO}}{\text{Im} \mathcal{F}^{LO}}=
\frac{t}{Q^2}c_t^{\mathcal{F}}(\xi,t)+
\frac{m^2}{Q^2}c_{m}^{\mathcal{F}}(\xi,t)\,,
\end{align}
where $\mathcal{F}=\{\mathcal{H},\mathcal{E}\}$,
\begin{align}
\text{Im}\, \mathcal{H}^{LO}=&\pi \sum_q e^2_q\big[H^q(-\xi,\xi,t)-H^q(\xi,\xi,t)\big]\,,
\end{align}
and similar for $\text{Im}\, \mathcal{E}^{LO}$. 
The coefficients $c_t^{\mathcal{F}}(\xi,t)$ depend on $t$ because of the non-factorizable
$t$-dependence of the GPDs through the Regge trajectory.

In Fig.~\ref{figure1} we show $c_t^{\mathcal{H}}(\xi,t)$ and $c_m^{\mathcal{H}}(\xi,t)$ as
a function of $\xi$ for several $t$-values: $-t = 0, 0.5, 1.0, 1.5$~GeV$^2$. The same is
shown in Fig.~\ref{figure2} for $c_t^{\mathcal{E}}(\xi,t)$ and $c_m^{\mathcal{E}}(\xi,t)$.
One sees that the corrections to $\mathcal{E}$ are in general larger than for the
$\mathcal{H}$ form factor; in particular $\mathcal{E}$ receives a relatively large proton
mass correction.

Finally, note that the finite-$t$ correction depends on the definition of the skewedness
parameter $\xi$, which is not unique. If one defines $\xi$ through the Bjorken $x_B$
parameter, $\xi_B= x_B/(2-x_B)$, which seems to be natural from the experimental point of
view (the relation of ``our'' $\xi$ to $\xi_B$ is given in Eq.~\!(125) in
Ref.~\cite{Braun:2012bg}), $c_t^{\mathcal{H},\mathcal{E}}(\xi,t)$ change accordingly, but
in general do not become smaller.

To summarize, in this work we have calculated, for the first time, the kinematic power
corrections $\sim t/Q^2$ and $\sim m^2/Q^2$ to the helicity amplitudes of deeply virtual
Compton scattering. These corrections are important for intermediate momentum transfers
$Q^2\sim 1-10$~GeV$^2$ that are accessible in the existing and planned experiments, and
have to be taken into account in the data analysis. In particular the finite--$t$
corrections are indispensable if one aims to study ``holographic'' images of the proton in
the transverse plane, in which case the $t$-dependence must be measured in a broad range.

\section*{Acknowledgements}
This work was supported by the DFG, grant BR2021/5-2.




%
%


\end{document}